\documentclass[twocolumn,twoside,slac_two]{revtex4}
\usepackage{graphicx}
\usepackage{fancyhdr}
\begin{document}

\title{Discovery of very-high-energy $\gamma$-ray emission from the vicinity of PSR J1831-952 with H.E.S.S.}

%

\author{F. Sheidaei}
\affiliation{Unit for Space Physics, North-West University, Potchefstroom 2520, South Africa, \\ Astroparticule et Cosmologie (APC), CNRS, Universit\'{e} Paris 7 Denis Diderot, 10, rue Alice Domon et L\'{e}onie Duquet, F-75205 Paris Cedex 13, France}
\author{A. Djannati-Ata\"{i}}
\affiliation{Astroparticule et Cosmologie (APC), CNRS, Universit\'{e} Paris 7 Denis Diderot, 10, rue Alice Domon et L\'{e}onie Duquet, F-75205 Paris Cedex 13, France}
\author{\& H. Gast}
\affiliation{Max-Planck-Institut f\"ur Kernphysik, P.O. Box 103980, D 69029 Heidelberg, Germany}
\author{For the HESS Collaboration}
\begin{abstract}
We report on the latest discovery of an extended Very High Energy  (VHE) $\gamma$-ray source near the 67 ms pulsar PSR J1831-0952 during the H.E.S.S. Galactic Plane Survey (GPS). The dispersion measure distance of the pulsar (4.3 kpc) would imply that less than $\sim1\%$ of its spin-down energy is required to provide the observed VHE luminosity of the source. No other plausible counterparts have yet been found through preliminary multi-wavelength searches. The most likely scenario is that the VHE emission originates from the -- yet unseen at other wavelengths --  wind nebula of PSR J1831-0952. If so  this would constitute another case of a $\gamma$-ray discovered pulsar wind nebula. 
\end{abstract}

\maketitle

\thispagestyle{fancy}


\section{Observations and Analysis}
The H.E.S.S. (High Energy Stereoscopic System, an array of four imaging atmospheric Cherenkov telescopes located in the Khomas Highland in Namibia) has revealed more than 60 sources of very-high-energy (VHE) $\gamma$-rays  through its Galactic Plane Survey (GPS) since 2004. 
Thanks to the use of advanced multivariate analysis techniques and the accumulation of exposure,  H.E.S.S. has  achieved a sensitivity of better than 2\% of Crab  in the core region of the GPS (i.e. $l = 282^\circ$ to $60^\circ $) \cite{Gast ICRC32}.
Pulsar Wind Nebulae (PWNe) constitute by far the dominating source population as compared to that  of young shell-type Supernova Remnants (SNRs), or to that of older and/or interacting remnants. About one third of H.E.S.S. sources have still either no known counterparts in other wavelengths, or lack any clear emission scenario. 
\\
The data on HESS J1831-098 consist of observations either dedicated to nearby sources such as SNR 21.5-0.9/HESS J1833-105, or from the extension of H.E.S.S GPS near $l=21^{\circ}$. These data were taken in 2004 (May-Oct.), 2005 (June and July), 2007 (Apr. and July), 2008 (Sept.) and 2009 (Apr.-July), for a total observation time on HESS J1831-098 of $\sim 52$ hours. After application of the H.E.S.S. standard data quality criteria \cite {aharonian2006a} based on hardware and weather conditions, the data set for HESS J1831-098 amounts to a total live-time of $40$ hours with average zenith angle and average offset (to the FoV center) of  $22.8^\circ$, and  $1.30^\circ$, respectively. The mean offset is rather large because observations were not specifically targeted at this source.
 \\
The standard Hillas H.E.S.S. event reconstruction scheme was applied to the data after calibration\cite{aharonian2004a}. The rejection of cosmic-ray showers was done by application of a recently developed multivariate analysis  \cite{Becherini}.  The sky maps were produced with an image size cut of 80 photo-electrons (p.e.) and using the {\it Ring Background} method \cite{Berge 2007} where the background at each test position on the sky is derived from a ring surrounding it with a mean radius of $0.7^\circ$.  To derive the spectrum, the same cut on image size was applied together with the a {\it Reflected-Region} procedure to estimate the Background,  followed by the application of a forward-folding method\cite{Piron 2001}.

\section{Results}
The excess count map of the $0.4^\circ \times 0.4^\circ$ region around HESS J1831-098 is shown on Figure 1. The map is smoothed with a Gaussian of $\sigma \sim 0.12^\circ$. An extended $\gamma$-ray emission to the south-east of PSR J1831-0952 is observed with a peak pre-trials significance of $ 7.9 \sigma$ for the standard integration radius of $\theta=0.22^\circ$, used for generation of the GPS maps when searching for extended sources. The significance level of the source after a conservative correction for trials is $5.8 \sigma$.   
The fit of the excess map with a two dimensional symmetrical Gaussian function, convolved with the H.E.S.S Point-Spread Function (PSF), results in a source centroid position of $\alpha\sim 18^{\rm h}31^{\rm m}25^{\rm s}, \delta \sim-9^{\circ}\,54^{'}$, with a width of  $\sigma \sim 0.15^\circ$ and a $\chi ^2$ of 593.4/525.
The fit of an asymmetrical Gaussian function does not improve significantly the $\chi^2$ nor the residuals. 
 \begin{figure}[!t]
  \vspace{1mm}
  \centering
\includegraphics[width=3 in]{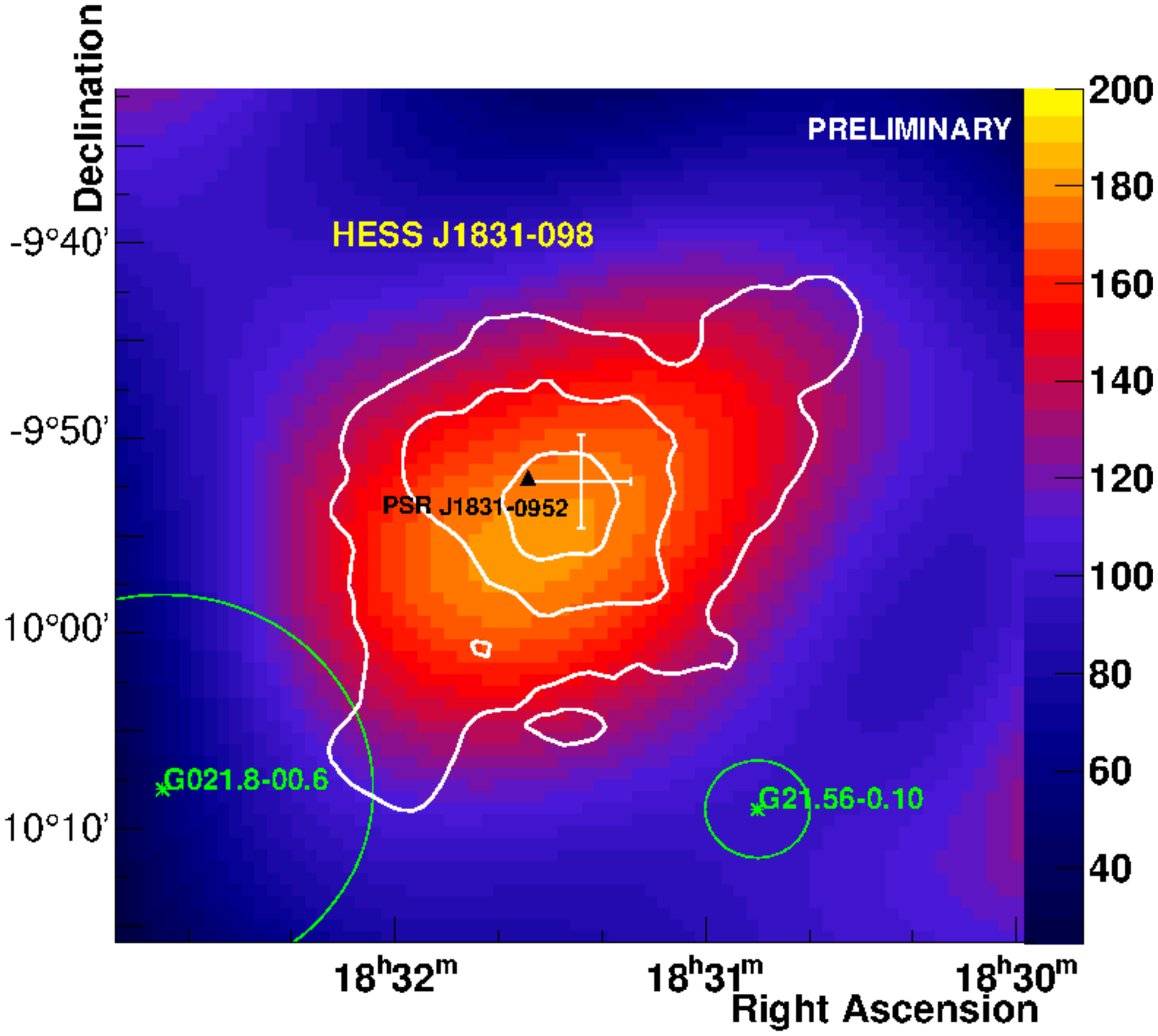}
  \caption{The excess map for HESS J1831-098, smoothed with a Gaussian of $\sigma \sim 0.12^{\circ}$. The white curves show the significance contours at 5, 6 and 7 $\sigma$ for an integration radius of $0.22^\circ$. The white cross shows the fitted position of the source. The position of PSR J1831-0952 is shown as a black triangle. Neighbouring SNRs are shown in green. Note that the fitted centroid does not coincide with the emission peak due to its departure from a Gaussian shape.} 
   \end{figure}
 \begin{figure}[!t]
  \vspace{5mm}
  \centering
  \end{figure}
\begin{figure}[!t]
 \vspace{5mm}
  \centering
\includegraphics[width=3 in]{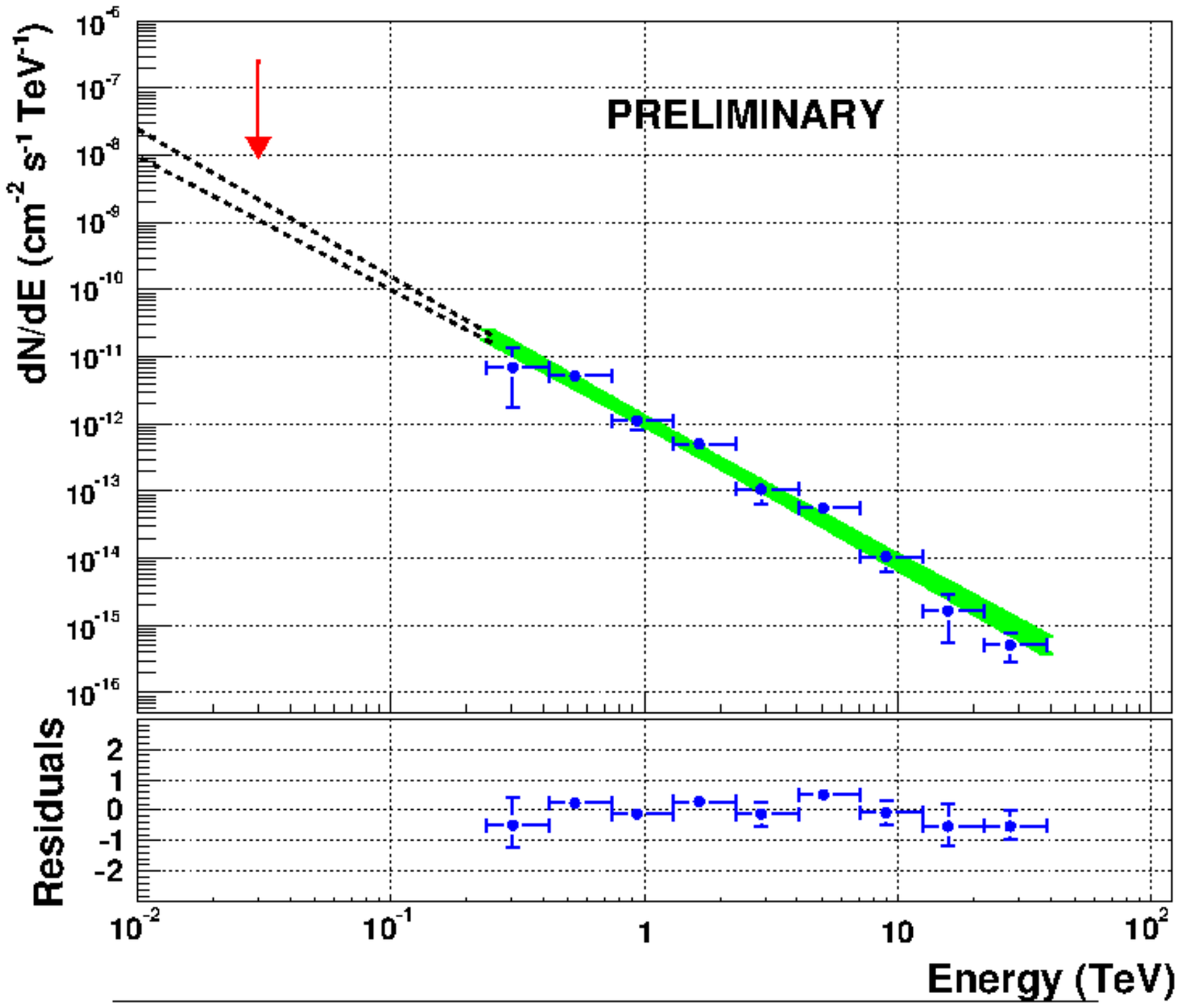}
 \caption{Differential energy spectrum of HESS J1831-098, fitted with a power-law function after extraction of events from a circular region of $0.3^\circ$ radius, centered on the best-fitted position. The arrow shows the Fermi differential U.L. at 30 GeV.}
 \end{figure}
 A circular region of $0.3^\circ$ radius, chosen as a compromise between optimal signal-to-noise ratio and independence of source morphology, was used to determine the energy spectrum. 
 The differential spectrum, based on the 484 excess events in the circular region, is shown in Fig. 2. The best fitted shape is a power-law ($\rm d \rm \phi/\rm{dE=\phi_{0} (E/1TeV)^{-\Gamma}}$) with no significant indication of a cut-off up within the fitting energy range which extends up to 30 TeV. The fitted flux (uncorrected for events falling outside the integration disk) is  $\rm{\phi_{0}= (1.1 \pm 0.1) \times 10^{-12} TeV^{-1}cm^{-2}s^{-1}}$ and the photon index  $\Gamma=2.1 \pm 0.1$. The integrated flux for $\rm{E>1\, TeV}$ corresponds to about 4\% of the flux of the Crab nebula in the same energy range \cite{aharonian2006a}. 
\section{Search for Counterparts}
A preliminary multi-wavelength search for energetic counterparts to HESS J1831-098 resulted in the sole pulsar PSR J1831-0952 \cite{Lorimer}, which lies at a small angular offset of  $\sim 0.05^\circ$  from the H.E.S.S. source's best fit position (see Fig. 1). Some X-ray data from {\it Chandra} and XMM-{\it Newton} are available, but the observations have been done at large offsets with respect to PSR J1831-0952 and hence are not very useful. 
The search in the GeV energy was carried out using $\sim 33$ months of public data from the Large Area Telescope (LAT) on board of Fermi (collected from 2008 August 4 (MJD 54682) to 2011 April 10 (MJD 55661)). A region of interest (ROI) of $6 ^\circ$ was used to select events in the [10-100] GeV range. The ROI radius was chosen such as to be large enough to get a reliable value for the normalization of the diffuse model and to be several times greater than the LAT PSF in the selected energy range. Events were analysed by applying the standard \textit{Fermi} Science Tools to events of class 4 which is recommended for studies of faint diffuse sources and which go beyond 20 GeV (in order to to minimize the non-photon background contamination). The corresponding instrument response function used for this event class is P6\_V3\_DATACLEAN \cite{Abdo 2009a}. Other standard cuts were also applied (e.g. zenith angle larger than $105 ^\circ$ in order to reduce the contribution from terrestrial albedo $\gamma$-rays \cite{Abdo 2009b}).
   
A hot-spot, at a pre-trials TS=25, was found within the boundaries of the H.E.S.S. source on a map generated by \textit{gttsmap} (Fig. 4). The hot-spot consists of one $0.01 \rm deg^2$ pixel at TS=25 and few neighbouring pixels at lower TS values. To account for the trials factor a conservative approach was adopted, namely  the ratio of the predefined HESS source circle of radius $0.15 ^\circ$ to one Fermi sky map pixel area was used. This yields a post-trial significance of $\sim 4.5 \sigma$ for the hot-spot.  Investigating a slightly higher energy band of  [30-100] GeV, the hot-spot significance drops to less than $3.4 \sigma$. Hence an upper limit (U.L.) on flux of $\rm{U.L. \sim 5 \times 10^{-11} ph cm^{-2} s^{-1}}$ was derived by assuming a spectral index equal to that of the HESS source ($\Gamma=2.1$). This U.L. is higher than the extrapolated flux of the H.E.S.S. source into the LAT range: $\rm{( [30-100]\, GeV) \sim 3.5 \times 10 ^{-11} ph \,cm^{-2} s^{-1}}$, and therefore does not exclude the existence of a GeV counterpart.   

\section{Discussion and Summary }
PSR J1831-0952 is an energetic pulsar exhibiting a spin-down luminosity of $\rm{1.1\times 10^{36} erg \,s^{-1}}$, spin-down age of $\rm{\tau_c \simeq 128}$ kyr, spin period of $\rm{\sim67\,ms}$ and a distance estimated from dispersion measurements of 4.32 kpc \cite{Manchester 2005}. To power the H.E.S.S. source, a conversion efficiency from rotational energy to 1-20 TeV $\gamma$-rays of $\epsilon \sim1 \%$ would be required, i.e. a value comparable to those inferred for other VHE PWN candidates such as HESS J1420-609/HESSJ1418-607 in the wings of Kookaburra \cite{Kookaburra}, HESS J1718-385 or HESS J1809-193 \cite{Aharonian 2007}. The angular offset of  $\sim 0.05^\circ$, translating into a projected distance of  $\rm{\sim 4 (\frac{d}{4 kpc})} \, \rm{pc}$, is also well within the range of offsets observed for crushed PWNe, as is the projected size of HESS J1831-098 ($\rm{\sim 20 (\frac{d}{4 kpc}) \, pc}$) when compared to the size of known VHE $\gamma$-ray PWNe.

These values and the fact that the offset-type morphology is rather common to VHE emitting PWNe (e.g. HESS J1825-137, MSH 15-52, HESS 1718-385 and HESS J1809-193) favour an interpretation of the VHE emission as originating from a PWN associated to PSR J1831-0952. In this scenario $\gamma$-rays are produced through Inverse Compton (IC) scattering of ambient  radiation fields (2.7 K Cosmic Microwave Background Radiation (CMBR), dust and star-light) by electrons injected by the pulsar into the nebula and re-accelerated at the wind terminal shock.  
The offset of the  VHE-peak from the pulsar position can be due to the expansion of the SNR/PWN into an inhomogeneous medium (e.g. \cite{Blondin}) and/or the proper motion of the pulsar. In the  latter case, and if the system age is indeed equal to the characteristic age of 128 kyr, the implied pulsar projected velocity of $\rm{\sim 300 \,km\,s^{-1}}$ remains reasonable and well within the bimodal velocity distribution derived in \cite{Arzoumanian}.

On the other hand, the extension of the maximum measured energy up to 30 TeV would imply a rather low magnetic field of ~1 $\mu \rm G$ in the IC scenario (see e.g., Eq. (6) in \cite{de Jager 2009}). However, the inferred value of the magnetic field intensity depends on the system age, which could suffer from large uncertainties (overestimation) given that the pulsar characteristic age is calculated assuming that the initial spin period is negligible and remarking this is not always the case (e.g. PSR J1400-6326, where the true age of 1-2 kyr, is much smaller than the characteristic age of 12.7 kyr \cite{Renauld}).
  \\
\begin{figure}[!t]
\vspace{7mm}
\centering
\includegraphics[width=2.5 in]{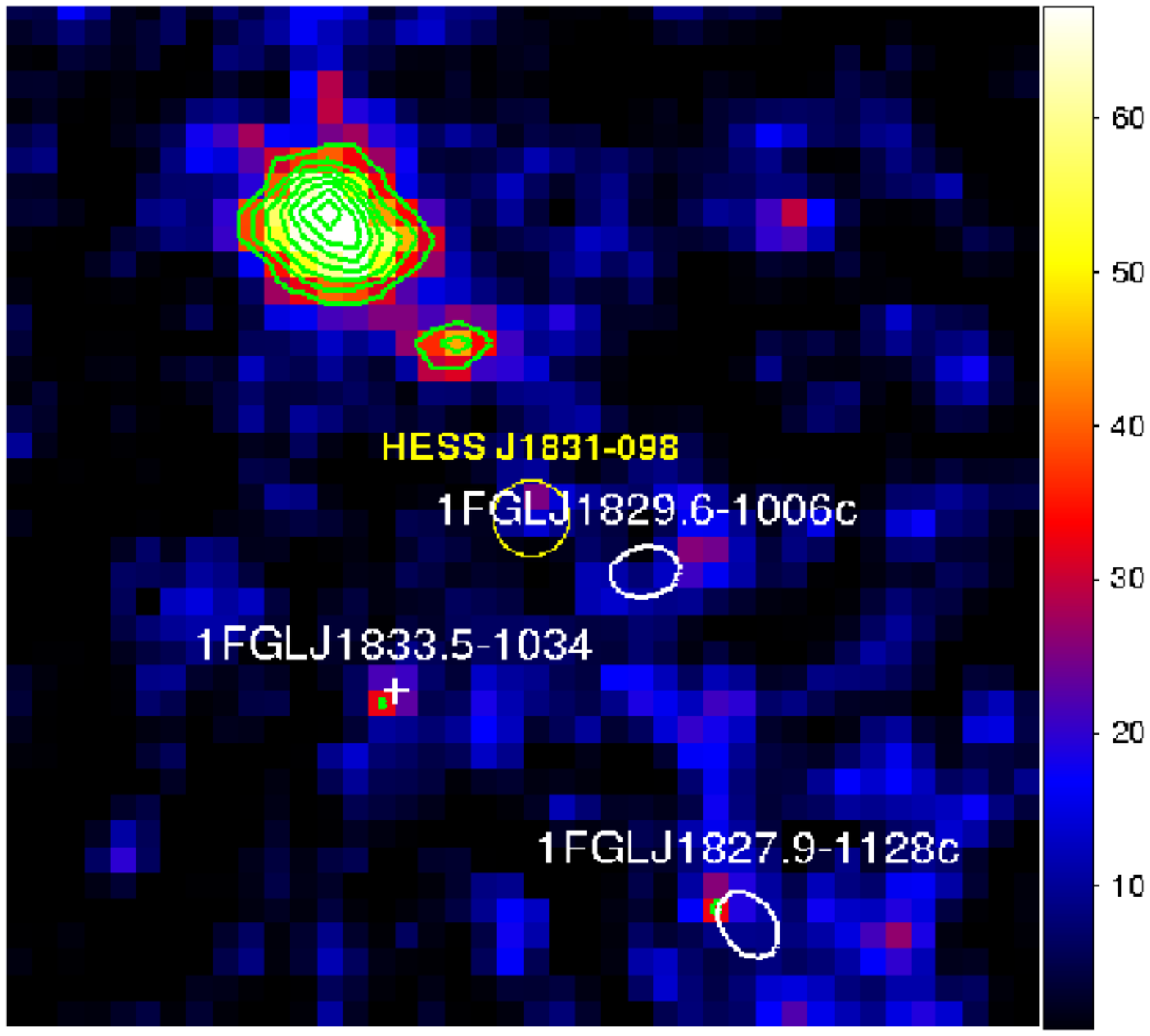}
\caption{Fermi TS Map in the 10-100 GeV range (FoV = $2 ^\circ$) with contours for $\rm{TS> 30}$ corresponding to $\sim 5.5 \sigma$ (pre-trials). HESS J1831-098 is shown by the yellow circle of radius $0.15 ^\circ$.}
\end{figure}

To summarize, although no PWN has been detected so far at other wavebands, the interpretation of  HESS J1831-098 in terms of a wind nebula remains the most likely scenario, given the spatial coincidence with the energetic PSR J1831-0952, the reasonable efficiency and offset, and the observed abundance of such PWN-type VHE sources.
If the association is confirmed,  HESS J1831-098 would constitute a gamma-ray discovered PWN. If so and if the true age of the system is close to the characteristic age of the pulsar, this source would be among the oldest known TeV PWNe. 
However, given the uncertainties on the source morphology due to limited statistics, other scenarios (e.g. a SNR shell emission) can not be excluded.  X-ray observations an additional data in the VHE $\gamma$-ray band are necessary to better understand the origin of the VHE emission.

\bigskip 
\begin{acknowledgments}
The support of the Namibian authorities and of the University of Namibia
in facilitating the construction and operation of H.E.S.S. is gratefully
acknowledged, as is the support by the German Ministry for Education and
Research (BMBF), the Max Planck Society, the French Ministry for Research,
the CNRS-IN2P3 and the Astroparticle Interdisciplinary Programme of the
CNRS, the U.K. Science and Technology Facilities Council (STFC),
the IPNP of the Charles University, the Polish Ministry of Science and 
Higher Education, the South African Department of
Science and Technology and National Research Foundation, and by the
University of Namibia. We appreciate the excellent work of the technical
support staff in Berlin, Durham, Hamburg, Heidelberg, Palaiseau, Paris,
Saclay, and in Namibia in the construction and operation of the
equipment.
\end{acknowledgments}

\end{document}